\documentclass{amsart}
\usepackage[mathcal]{euscript}
\usepackage{amssymb, amsfonts, xypic}
\usepackage{graphicx}
\usepackage{float}
\floatstyle{boxed}
\restylefloat{figure}

\begin{document}
\title{Simulating Factorization with a Quantum Computer}
\maketitle
\begin{center}
\author{Jose Luis Rosales}

\footnotesize jose.rosales@fi.upm.es

\itshape 

Center for Computational Simulation,

DLSIIS ETS Ingenieros Inform\'{a}ticos, Universidad Polit\'{e}cnica de Madrid

Campus Montegancedo, E28660 Madrid

\end{center} 
\normalfont

\begin{abstract}
%Quantum entangled tokens could be an option to classical asymmetric  network security based on prime factorization cryptosystems. Thus, in order to decide whether current  security is strong, 
Modern cryptography is largely based on complexity assumptions, for example, the ubiquitous RSA is based on the supposed  complexity of the prime factorization problem. 
Thus, it is of fundamental importance  to understand how a quantum computer would eventually weaken these algorithms.  In this paper, one follows Feynman's prescription for a  computer to simulate the physics  corresponding to the algorithm of factoring a large number $N$ into primes. Using Dirac transformation theory one translates factorization into the language of  hermitical operators, acting on the vectors of the Hilbert space. This leads to obtaining the ensemble of factorization of $N$ in terms of the Euler function $\varphi(N)$, that is quantized.  On the other hand, a quantum  mechanical prime counting function $\pi_{QM}(x)$, where $x$ factorizes $N$, is derived. This function converges to $\pi(x)$ when $N\gg x$.  It has  no counterpart in analytic number theory and its derivation relies on semiclassical quantization alone.
\end{abstract}
\footnotesize
I dedicate this paper to Professor Emeritus Jos\'{e} Luis S\'{a}nchez-G\'{o}mez.   
\normalsize
\section{Introduction}
\label{sec:Introduction}

 The computational effort  required to find  the prime factor of a large number $N$ is the basis for the widely used RSA public key encryption system. This makes relevant the study of new methods and algorithms of factorization. The procedures of all classical computers algorithms perform different kind of sieves\cite{sieves} of the primes $\{x\}\leq N^{1/2}$, but, despite the many mathematical successes, the factorization time  of all classical algorithms still scales, at most, sub exponentially in the number of bits of $N=2^b$. 

%if $x$ and $y$ are the factor of some integer $N$, then factoring $N=xy$ is equivalent to obtaining the  Euler's  function $\varphi(N)=N(1-1/x)(1-y)=(x-1)(y-1)$,  that provides all the integers relatively primes  to $N$. This is because recall that $\varphi(N)$ could also be seen as the discrete Fourier transform of the function $gcd(k,N)$ for all integers $k\leq N$,\[\varphi(N)=\sum_{k=1}^{k=N}gcd(k,N)\exp(-2\pi i k/N).\]\\ 
Suppose one finds,  for some integer $a$,  $gcd(a,N)=1$, the period $r(a;N)$ of the function $f(a;z)=a^{z}\ mod\  N,$ $ =f(a;z+r)\ mod\  N,$  then one gets the factors $N=x_{+}x_{-}$ simply as
\[x_{\pm}=gcd(a^{r/2\pm 1},N).\]\\ 
Shor has found  an algorithm that implements the quantum Fourier transform of $f(z)$ yielding to the value of $r$ in polynomial time in a quantum computer \cite{shor}. Yet, the current state of the art of quantum computation would not allow for  factorization if $N$ is large.

The grounds  of  quantum computation  speed up versus classical computation is  the property of entanglement of different states amplitudes (linear superposition) for the output of physical systems (qubits) that represents the logic of the algorithm itself.
Moreover,  the very principle of quantum computation is Feynman's observation,  describing the problem of simulating quantum physics with computers\cite{Feynman} which, in the end, is equivalent to obtaining a pseudo-probability measure from the unitary evolution of Wigner's function,  calculated for the input states of the computer. Generally speaking, it would involve the whole Hilbert space of the physical system we intend to simulate, and this is the profound difficulty to perform  real quantum computations.

Now, following Feynman's observation, we could ask for the possibility to obtaining the Hilbert space of some (already) quantum computer that performs factorization. This is the approach of the paper, in this case, obtaining the physics from the classical algorithm (conversely to standard quantum computation that requires the realization of a physical system  performing the algorithm).  These are the grounds and methods of theoretical  physics arising from semiclassical quantization of  arithmetical functions once they are translated to the language of Hamiltonian mechanics. To this aim, we will transform the number theoretical functional  of factorization, described  in section  2, precisely into the Jacobi functional defined in the computer coordinates space. It  prescribes a methodology  based on Dirac-Jordan transformation theory\cite{dirac} and Feynman's class of quantum computers (full Hilbert space) in number theory.  

Moreover, this could also be consistent with the hypothesis of Hilbert and P\'{o}lya (see \cite{Mo}) related to, more precisely,  defending the idea that the the Riemann function $\zeta(s)$ zeros, should be obtained from the spectrum of a suitable Hermitical operator (the Feynman Quantum Computer for Riemann's $\zeta(s)$) . It amounts to affirm the Riemann hypothesis\cite{ri}, i.e., for such an Hermitical operator,$H$, the Schr\"{o}dinger equation, \[H\psi_{m}(q)=T_{m}\psi_{m}(q)\]\\holds in some parametric unbounded space ${q}$ acting on the Hilbert space vectors $\psi_{m}$ and the (real) eigenvalues  $T_{m}$ should necessarily correspond to the zeros of $\zeta(s)$ on the critical line $s=1/2+iT_{m}$.

The meaning of the Hilbert space $\{\psi_{m}(q)\}$, parametric space $\{q\}$ and the quantum conditions remains an open question. Some ad hoc ideas dealing with such a Hilbert construction have been proposed so far, the more promising being that of Berry and Keating \cite{B-K} relating the semiclassical quantization of some yet abstract physical system with the properties of Riemann $\zeta(s)$. In this work we will not follow on these investigations directly, rather we will derive a classical Hamiltonian and the Hilbert construction for the factor $x$ of $N$; this  is indirectly related to the theory of primes upon its dependency with the theory of factorization of a number $N$. 

Therefore, similarly to those ideas, we intend to directly calculate from the methods of semiclassical quantization the exact values of such a well defined arithmetical function. The Hamiltonian methods and quantum transformation theory are the resources needed to obtain: (1) the dimensionality of the Hilbert space, (2) the semi classical parametric space $\{q\}$ and (3) the  unitary state basis $\{\psi_{m}\}$. In the end, quantization amounts to be equivalent to the requirement of divisibility and, in this sense, the discreteness of the factors of an integer $N$  satisfying the limits derived from the definition of the \itshape {classical} \normalfont Hamiltonian  is predicted and modeled. As in standard quantum mechanics, the correspondence principle is required to recover classical number theory. This is modeled upon the detailed calculation of the constants in the integral of the solutions of the quantum conditions.    

Our results are those of a Sturn-Liouville problem in quantum mechanics obtaining the factorization ensemble. It leads to  a   methodology to exhaust all the prime factor candidates of $N$.   A proof of the successfulness  of the present approach is the arrival to a new expression for the prime counting function  $\pi_{QM}(x;N)$, derived from the exact quantum mechanical solution,  having no counterpart in number theory, depending on $N$ as a parameter, if $x$ factorizes $N$; the exactitude of this quantum mechanical prime counting function increases indefinitely for large $N$ and it will eventually become a better approximation than $\pi(x)\sim Li(x)$. Indeed, \[ \pi(x)-\pi_{QM}(x;N)\rightarrow 0\]\\ for $N\gg 1$. This will be the main result of this paper.  

If RSA security can be weakened in this way by a quantum computer, one would be led to consider quantum tokens as the alternative to classical network security\cite{shor2}.

\section{Prime factorization  and Hamiltonian Mechanics}
\label{sec:PrimeFactorizationAndHamiltonianMechanics}

To simulate the computer of factorization we first require the classical theory. In physics what one uses is the Jacobi functional of the system written in canonical coordinates. So, firstly we are obliged to write a kind of functional analogous law to Jacobi's having the meaning of Euclides's algorithm for prime factoring $N$ 
\[   \pi(x)\leq \pi(N^{1/2})\]
for the primes candidates $x$.
 
Now, in order to find how the algorithm works in the computer, we should find how this restricts the set of the $x$ for some input $N$. Recall now that for the input $N$ there are many other $N_{\sigma}$, close enough to $N$ that satisfy $j=\pi(N^{1/2})=\pi(N_{\sigma}^{1/2})$, so $x$ can be a factor of all those $N_{\sigma}$'s in a neighborhood $\mathcal{F}(j)$ of the input $N$. Define now, for $N_{\sigma}\in \mathcal{F}(j)$, 

\begin{equation}
	E\{j;x_{\sigma}\}=\frac{\pi(x_{\sigma})\pi(N_{\sigma}/x_{\sigma})}{j^2}.
\end{equation}\;

And $\mathcal{F}(j)=\{N_{\sigma}$; $\pi(N_{\sigma}^{1/2})=j \}$; $x_{\sigma}\cdot y_{\sigma}=N_{\sigma}\in\mathcal{F}(j)$, for the primes $(x_{\sigma}, y_{\sigma})$, and  $E_{\sigma}$  in Eq. (1)  defined in $\mathcal{F}(j)\subset \Pi\{2\} $ ( $\Pi\{2\} $ being the set of all integers having two prime factors).

For some input number $N$, $\mathcal{F}(j)$ would be the factorization ensemble of $N$. Now if for $N_{\sigma}$ integer, we could find another prime  $y_{\sigma}=N_{\sigma}/x_{\sigma}$ such that equation (1)  obtains a  single value of $E_{\sigma}$, then $\mathcal{F}(j)$ defines a proximity of $N$ that exhausts all the prime factor candidates pairs $(x_\sigma, y_\sigma)$ of $N$.

Moreover, given that, by definition, $N\in \mathcal{F}(j)$, counting all the $N_{\sigma}$ in $\mathcal{F}(j)$, i.e., finding the cardinality  $F(j)$, will obtain the algorithmic complexity of factoring $N$.\;

Prime Number Theorem\footnotemark[2]\footnotetext[2]{I thank to F.A. Gonzalez-Lahoz for these insights (private communication).}   obtains how  $F(j)=Card\{\mathcal{F}\}$  scales for large N. Let $\pi(N^{1/2})=j$, then, since $\mathcal{F}(j)\subset \Pi\{2\} $,
\begin{equation}
F(j) =\sum_{\pi_{2}(j-1)}^{\pi_{2}(j)} 1= \sum_{i=1}^{j-1}[\pi(x(j)^2/x_{i})-\pi(x(j-1)^2/x_{i})]=f(j+1)-f(j-1)+j,
\end{equation}\;
where \[f(j)=\sum_{i=1}^{j-1}\pi(x(j)^2/x_{i})\] can be computed asymptotically as 
\[\rightarrow\sum_{p=2}^{x(j)}\frac{x(j)^2}{p (\ln(x(j)^2/p)-1)}=\frac{x(j)^2}{\ln(x(j)^2)-1}\{\sum_{p=2}^{x(j)}\frac{1}{p}+\frac{\ln p}{p(\ln(x(j)^2/p)-1)}\}
,\]
now take $x(j)\approx N^{1/2}$; after some straightforward albeit long calculations, replacing the sums by integrals \[\sum_{p}^{x(j)} f(p)\simeq\int _{2}^{x(j)}\frac{f(y)}{\ln y} dy +\cdots,\]\;
one gets
\begin{equation}
F(j)\simeq  N^{1/2}\cdot(\ln\ln N^{1/2}+C)\cdot\{1-\ln\ln N^{1/2}/\ln N^{1/2}+1/\ln N^{1/2}+\cdots.\}
\end{equation}\;

Here $C(N\gg 1)\simeq \ln 2+B_{1},$ where \[B_{1}=\lim_{L\rightarrow \infty} \sum^{L} 1/p-\ln\ln (L)=0.2614972\cdots\]\\ is Meissel Mertens constant.\;

Since finding the co-prime $y_{\sigma}$ imposes  $\pi(x_{\sigma})\leq \pi(N^{1/2})$ one would expect statistically,  as many as $ N^{1/2}/x_{\sigma}$possible values of $y_{\sigma}$ per each $x_{\sigma}$ in $\mathcal{F}(j)$.\\

Hence, we could replace the problem of finding the prime factor $x$ of  $N$ with that of solving  the more general functional equation

\begin{equation}
E\{j;x_{\sigma}\}=E(N,x)).
\end{equation}\;

Now rework Eq. (4) introducing the variables $p$ and $q$.
\begin{center} 
\begin{equation}
p=\frac{\pi(y_{\sigma})-\pi(x_{\sigma})}{2 j}
\end{equation}\;
\begin{equation}
	q=\frac{\pi(y_{\sigma})+\pi(x_{\sigma})}{2 j}.
\end{equation}\;

\end{center}

We recast Eq. (4) in the following suggestive way,
%\begin{center}
\begin{equation}
-p^2+q^2=E
\end{equation}\;
%\end{center}

Whose solution is that of a classical inverted harmonic oscillator \[p=E^{1/2}\sinh(t),\]\\ \[q=E^{1/2}\cosh(t).\]\\ For $N>>1$ the functional $E$ is essentially a step function (because for the same value of $x_{\sigma}$ there are about $\sim  N^{1/2}/x_{\sigma}$ values of $y_{\sigma}$ with almost the same  value of $E$). 

Along with the computation of $E$ from Eq.(7), we might have considered  variations in $p$ and $q$ due entirely to changes in $t$. Now  $t$ is approximately a quasi-continuum parameter at $x<< N^{1/2}$ and, at those conditions, it has the meaning of the time variable in Hamilton's equations ($E$ an adiabatic invariant  in the variation)
\begin{equation}
\delta p=-\partial_q H\delta t,
\end{equation}\;

\begin{equation}
\delta q=\partial_p H \delta t
\end{equation}\;

$H$ being the Hamiltonian of the canonical coordinates $p$ and $q$.
\begin{equation}	
	H(p,q)=\frac{1}{2}(-p^2+q^2).
\end{equation}\;
 
Moreover $p=\partial_q S(q)$, so that a Hamilton-Jacobi condition exists for the functional $S(q)$

\begin{equation}
H(\partial_q S(q),q)=E/2;
\end{equation}\;

Jacobi's functional $S(q)$ is the analogous to the number-theoretical functional $E(x')$.

Equation (11) is relevant because $q$ is bounded and therefore its solutions are confined trajectories in parametric space.
\begin{equation}
q\leq\frac{\pi(N/2)+\pi(2)}{2 \cdot \pi(N^{1/2})}\sim \frac{1}{8} N^{1/2}. 
\end{equation}\;

$\omega=\partial_S H$ would be related to the quasi-period of those confined trajectories and we are led now to the conditions of semiclassical quantization. 

\section{Quantization}
\label{sec:Quantization}

In the $E$ representation, consider the state $\psi_{E}(q)$, that determines the quantum amplitude of probability of a  system semiclassically picked precisely on the classical trajectory given by $S(q)$.   

The state of the computer is
\[\Psi(q;j)=\sum_{k=1}^{k=F(j)}\frac{1}{F(j)^{1/2} }\psi_{E_{k}}(q)\]

Where the dimension of the Hilbert space of the computer that factorizes $N$ is assumed to be $F(j)$.

Now, the Hamilton-Jacobi constraint for $S(q)$ and quantum transformation theory allow us to obtain the momentum  operator acting on the wave functional $\psi_{E}(q)$ for  the q-numbers.
\begin{equation}
p\rightarrow-i\partial_q
\end{equation}\;
 
The Hamiltonian constraint in Eq. (7) becoming  a Hermitical operator in our canonical coordinates acting on $\psi$. \;\footnotemark[3] 
\footnotetext[3]{This is similar to what  Berry and Keating \cite{B-K} did while searching  the distribution of Riemann zeros, their conjecture supporting Hilbert and P\'{o}lya hypothesis concerned on the existence of some Hermitical operator whose eigenvalues $T_{k}$ correspond to Riemann zeros: $\varsigma(s)=0$ , $s=1/2+iT_{k}$.}

\begin{equation}
\frac{d^2 \psi(q)}{dq^2}+q^2\psi(q)=E \psi(q),
\end{equation}\;
\\

Our coordinate space satisfies  $E\leq q \leq\ qm(N)$, where $qm(N)< N^{1/2}/8$, therefore, our quantum conditions should be
\begin{equation}
\psi(E^{1/2})=0 ; 
\end{equation}\;
\begin{equation}
\psi( qm(N))= 0.
\end{equation}\;

The dimension of the Hilbert space of $\psi$ is the cardinality of the  factorization ensemble, $I\{\mathcal{F}(j)\}$.\footnotemark[4]\footnotetext[4]{ If we were to use the Berry and Keating quatization we should use a canonical transformation of our coordinates $q^{'}= p+q$ and $p^{'}=q-p$. Notwithstanding with the fact that  Berry-Keating classical Hamiltonian is simpler  than ours, $H_{BK}=1/2 p'\cdot q'$, its quantization requirements must encompass  quantum transformation theory along with a factor-ordering arbitrarity  to get an Hermitian operator \[H_{BK}\rightarrow-i/2(\frac{1}{2}+q^{'}\cdot\partial_q^{'})\]\\ acting on  $\psi$.}       

The Schr\"{o}dinger equation (14) and the Sturn-Liouville constraints  (15) and (16) define the eigenvalue problem leading to the quantization of $E$ without further assumptions. In this sense, dicreteness of the prime factorization of $N$ is a natural consequence of quantization.

In order to solve equations (14), (15) and (16) one makes $\rho=q^2$ and $\psi=\phi/\rho^{1/4}$, obtaining
\begin{equation}
\frac{d^2}{d\rho^2}\phi+\frac{l(l+1)}{\rho^2}\phi+2\mu(r^2-\frac{z^2}{\rho})\phi=0,
\end{equation}\;
where $l=-1/4$, $\mu=1/2$, $r=1/2$ and $z^2=E/4$.

That is the 3-dimensional Schr\"{o}dinger equation for the coulombian scattering of two identical charged particles in their center of mass. Quantum theoretically, the spectrum of $E$ corresponds  to the quantization of electricity of some system under the conditions of confinement in (17)\footnotemark[5].  \footnotetext[5]{Recall that Bhaduri et al.\cite{bal} postulated  the same equation of the inverted harmonic oscillator to reach the spectrum of Riemann zeros in exactly the same spirit than Berry-Keating\cite{B-K}; in our case, though, we derived the quantum conditions; even though we are rather concerned with the distribution of  prime factors of an integer $N$, it is a satisfactory coincidence that the quantum Hamiltonian we derived is just the same.}  

The solution of (17) is asymptotically for $\rho\gg 1$

\begin{equation}
\phi(\rho)\sim \sin(\frac{\rho}{2}-\frac{E}{4}\ln(\rho)+\delta_{Coul}(E)+\delta_{0}(E)),
\end{equation}\;
where $\delta_{Coul}(E)=Arg\{\Gamma(-i\frac{E}{4}+\frac{3}{4})\}$ and 
\begin{equation}
\delta_{0}(E)=-A\:N^{1/2}\!\ln(E)-h_{1}+o(1/E^2);
\end{equation}\;

$A$ and $h_{1}$ being the two (yet arbitrary) integration constants of our second order differential equation; $\delta_{Coul}(E)$ is a shift in the distorted Coulomb wave for the asymptote while $\delta_{0}$ represents the additional phase drift obtained  from the first condition at $\rho=E$, $\phi(E)=0$ using the general solution of (17) \cite{lan}
\begin{equation}
\phi(\rho)=\rho^{3/4} e^{-i\rho/2}\{U(\alpha,3/2,i\rho)+\cot\delta_{0}(E)F(\alpha,3/2,i\rho)\}.
\end{equation}\;
Here $F(a,b,c)$ and $U(a,b,c)$ are the confluent hypergeometric functions, $\alpha=-i\frac{E}{4}+\frac{3}{4}$. 

Now if $\rho\gg 1$ \[ F(\alpha,1/2,i\rho)\sim \frac{1}{\Gamma(\alpha)}(i\rho)^{\alpha}\exp(i\rho),\]\\ and \[U(\alpha,1/2,i\rho)\rightarrow 0,\]\\ 
which directly obtains  Eq. (18). Moreover, the logarithmic behavior in $\delta_{0}$ results also from these aymptotics and the fact that, \[U(\alpha,1/2,iE)\simeq (iE)^{-\alpha}= \exp(i\pi\alpha/2)(1+\ln(E)\alpha+\cdots ).\] The condition $\phi(E)=0$, yields to Eq. (19).

Eq.(19) is the first quantum condition for the confined wave, the second  $\phi(N/64)=0$  imposes from (18)
\begin{equation}
\delta_{Coul}(E)+\delta_{0}(E)+N/128-E/4\ln(N/64)=n\pi
\end{equation}\;
where $n$ is an integer number.

Redefining  $n$ \footnotemark[6] \footnotetext[6]{ We will take the convention that large $k$'s map the region $E\gg 1$ or  $\pi(x_{\sigma})\ll \pi(N^{1/2})$ also small quantum numbers correspond to $\pi(x_{\sigma})\approx \pi(N^{1/2})$ and in this case $E\rightarrow 1$.}, 
\begin{equation}
n=\left[N/(128\pi)\right]-k,
\end{equation}\; 
$k$ an integer, $1\leq k\leq F(j)$, also taken into account that for large $E$

\begin{equation}
\delta_{Coul}(E)=Arg\{\Gamma(-i\frac{E}{4}+\frac{3}{4})\}\sim E/4(1-\ln(E/4)+o(1/E^2)).
\end{equation}\;

Define now, for convenience in the notations, \[\gamma= j/N^{\frac{1}{2}}\ll 1. \]\.
Thus equation (21) leads asymptotically  $N\gg 1$ to
\begin{equation}
\frac{N^{1/2}\gamma}{\pi}\ln(E)\{1-\frac{E(\ln2-1/2)}{2A N^{1/2}}\}+\frac{E}{2\pi N^{1/2}}(1-\frac{3}{2}\gamma)= \frac{\gamma}{A}(k-h_{1}/\pi)+ o(1/E^2)
\end{equation}\;
%\begin{equation}
\[\rightarrow \frac{N^{1/2}}{\pi}\ln(E)=\frac{1}{A}(k-h_{1}/\pi)+ o(1/E^2).\]\\
%\end{equation}\;

\normalsize Eq. (24) is the second quantum condition and represents the quantization of $E$. It's just Bohr-Sommerfeld quantization of the states $\psi_{k}(E;q)$.\\ 
	
Let us see how $E$ scales for $x_{\sigma}\ll  N^{1/2}$. The functional of factorization attains its maximum at $\pi(3)=2$,

\begin{equation}
 E(x_{\sigma}=3)= 2\frac{\pi(N/3)}{j^2}\simeq \frac{1}{3\gamma},
\end{equation}\;
for $N\gg 1$. 
Eq. (25) suggests the Ansatz for the Quantum Mechanical Asymptote of the arithmetical functional $E$
\begin{equation}
E_{QM}=E_{1}\cdot\gamma^{\kappa_{1}-\kappa};
\end{equation} \;
where 
\[ \kappa\equiv k/I \leq 1\sim O(1),\] 
\[ \kappa_{1}\equiv k_{1}/I\]  
and $E_{1}=O(1)$ is a constant related to $k_{1}$. 

Feeding this back into Eq. (24) one directly obtains the appropriate values of $A$ and $h_{1}$: 
\[A\simeq \frac{\pi I}{N^{1/2}\ln(1/\gamma)}=O(1),\]\;

and  
\[h_{1}\simeq \pi  k_{1}.\]\;

Therefore, remarkably, the Ansatz (26) becomes the exact asymptotic solution of (24). 

Yet, $k_{1}$ must be determined from consistency with the PNT asymptote of $E$ at large $N$\footnotemark[8].

\footnotetext[8]{ Recall that $E_{\sigma}$ is a step function, i.e., for instance, it takes almost the same value $E_{}\sim1/3\gamma$ when $k$ belongs to the interval
\begin{equation}
k\in[I-\frac{ 5 N^{1/2} }{6},I- \frac{N^{1/2}}{2} ).
\end{equation}\; 
i.e., for those $k's$ representing the co-primes $y_{\sigma}$ of $N_{\sigma}\in \mathcal{F}(j)$ with $x_{\sigma}=3$ 

Thus, in the upper limit we subtracted to $F(j)$ the co-primes with $x_{\sigma}=2$, i.e. $N^{1/2}/2$ while, in the lower limit, we did the same for those co-primes corresponding to $x_{\sigma}=3$, namely $N^{1/2}/2+N^{1/2}/3$. This exhaust all the possible values of $k$ that attains $E_{3}$.

Therefore, at $\kappa(3)=k_{\small{min}}(3)/I$, Eqs.(25)  and (26) taken into account, a relation between $\kappa_{1}$ and $E_{1}$ holds 
\begin{equation}
E_{1}=1/3\gamma^{-(5/6 N^{\frac{1}{2}}/F(j)+\kappa_{1})}.
\end{equation}\; }

Now, in Eq. (1) put $\pi(x)\sim Li(x)$; etc.,  to obtain another asymptote for $E(x;N)$ if $x=O(N^{1/2})$    
\begin{equation}
E_{Li}(N; x)\simeq (1-u(x;N)^2)^{-1}
\end{equation}\; 
With the help of the arithmetical function  \[u(x;N)=\gamma\ln(N^{\frac{1}{2}}/x)\]\\ defined for the prime factor $x$ of $N$. 

Moreover, $u$ is related to Euler's function 
\begin{equation}
\varphi(N_{\sigma})=(x_{\sigma}-1)\cdot(N_{\sigma}/x_{\sigma}-1)=N_{\sigma}-2\!N_{\sigma}^{1/2}\cosh(u/\gamma)+1.
\end{equation}\;

Now, for $N_{\sigma}\in\mathcal{F}(j)$, $E=f(u)$  is quantized and so are the functions of $u$.Comparing Eqs. (29) and (26),  it predicts the existence of  a map \[u(x;N)\leftrightarrow \kappa(j).\] 

Moreover, the function 
\[( N_{\sigma}^{1/2}-\frac{\varphi(N_{\sigma})}{N_{\sigma}}+N_{\sigma}^{-1/2})=\cosh(u/\gamma),\]\\  takes  discrete values in $\mathcal{F}(j)$.   

Technically, in order to derive explicitly $u(\kappa)$, one has to find and solving the Schr\"{o}dinger equation for $\psi_{\varphi}(q_{\varphi})$, provided  the canonical transformations
\[p_{E},q_{E}\rightarrow p_{\varphi}, q_{\varphi},\]

Instead of doing this, let's follow a  straightforward approach to obtaining $u(\kappa)$  simply  using the known statistics for $N_{\sigma}\in\mathcal{F}(j)$. This can be done upon calculating a minimal (three points) Lagrange Polynomial fit between $\kappa$ and $u$ in our region of interest $u\sim O(1)$ where the approximation $1/E^2$ is valid. This amounts to selecting

\[u(3)\approx 1 \leftrightarrow \kappa(3)=1-\frac{5/6N^{1/2}}{F(j)},\] 
\[u(N^{1/2})=0\leftrightarrow\kappa(N^{1/2})=0\]
and 
\[u (2)=\gamma \ln(\frac{ N^{1/2}}{2})\leftrightarrow \kappa(2)=1.\]

This being done, it provides asymptotically
  
\begin{equation}
u(\kappa)= \alpha(N)\kappa -\beta(N)\kappa^2
\end{equation}\;
where $\alpha$ and $\beta$ are
\begin{equation}
\alpha(N)\simeq 2+O(\gamma\ln 1/\gamma );
%u(2)+\frac{1/\kappa(3)-u(2)}{1-\kappa(3)}
\end{equation}\;
\begin{equation}
\beta(N)\simeq 1+O(\gamma\ln 1/\gamma).
%\frac{1/\kappa(3)-u(2)}{1-\kappa(3)}
\end{equation}\;

Then,
\begin{equation}
\kappa\simeq \alpha/(2\beta)- 1/\beta (\alpha^2/4-\beta u)^{1/2}
\end{equation}\;

Selecting $E_{1}$ such that $E_{QM}(\kappa_{1})=E_{Li}(u(\kappa_{1}))$ obtains :
\begin{equation}
  E_{1}(\kappa_{1})\simeq (1-u(\kappa_{1})^2)^{-1}.
\end{equation}

This taken into account, Eq.(26) finally yields to\;  
\begin{equation}
E_{QM}(x;N)=  E_{1}(\kappa_{1})\cdot \gamma^{\{1/\beta (\alpha^2/4-\beta u(x;N))^{1/2}+\kappa_{1}-\alpha/(2\beta)\}}\;
\end{equation}

\;Here $\kappa_{1}$, after Eqs. (28) and (35) is the series\;   
\begin{equation}
\kappa_{1}\simeq (1/6)/\ln (1/\gamma)+R/\ln(1/\gamma)^2+ O(1/\ln(1/\gamma)^3)\},
\end{equation}\;
\;and $R= (5/6)C$ is a numerical constant.\; 

\section{Prime counting function $\pi_{QM}(x;N)$ }
\label{sec:Primecountingfunction}

Even though (36) is just the solution for $\ln x\ll 1/2\ln  N$, by construction it would become  exact in its range of validity ($\kappa\geq \kappa_{1}$, say).Then, we might use it to  obtain a completely new approximation for the prime counting function $\pi(x)$, i.e., for \[x< N^{1/2}\exp(-u(\kappa_{1})/\gamma).\]

\begin{equation}
\pi_{QM}(x;N)\equiv E_{QM}(x;N)\frac{\pi(N^{1/2})^2}{\pi(N/x)}
\end{equation}

Recall that $E(x;N)$ can be rewritten simply as
\[E(x;N)=x\frac{\pi(N/x)}{N\gamma}\cdot \frac{\pi(x)}{x\gamma}\]
and since PNT asymptote yields to
\[x\frac{\pi(N/x)}{N\gamma}\simeq\frac{1}{1+u},\]
Eq. (38) obtains 
\[ \pi_{QM}(x;N)\rightarrow E_{1} \gamma x\;(1+u)E_{QM}(x;N);\]\\
%\exp\{-\ln (1/\gamma) \cdot(-\alpha/(2\beta)+1/\beta(\alpha^2/4-\beta u)^{1/2}+\kappa_{1}})\},\]
where $u\simeq 1-\gamma(\ln x -1).$
%\[\rightarrow E_{1} \gamma x(1+u)\gamma^{\{1/\beta (\alpha^2/4-\beta u)^{1/2}+\kappa_{1}-\alpha/(2\beta)\}}\]\\
 %\[u\simeq 1-\gamma\ln(xe/8)\]\\ 

$\gamma$, $E_{1}$ and $\kappa_{1}$ being known functions of $N$. To derive the explicit formula above we took $\pi(N/x)\sim Li(N/x)$ and $\pi(N^{1/2})\sim Li(N^{1/2})$, for $N\gg 1$.The larger $N$ in (38) the better its exactitude.

\begin{figure}[hbtp]
\centering
\includegraphics[scale=0.4]{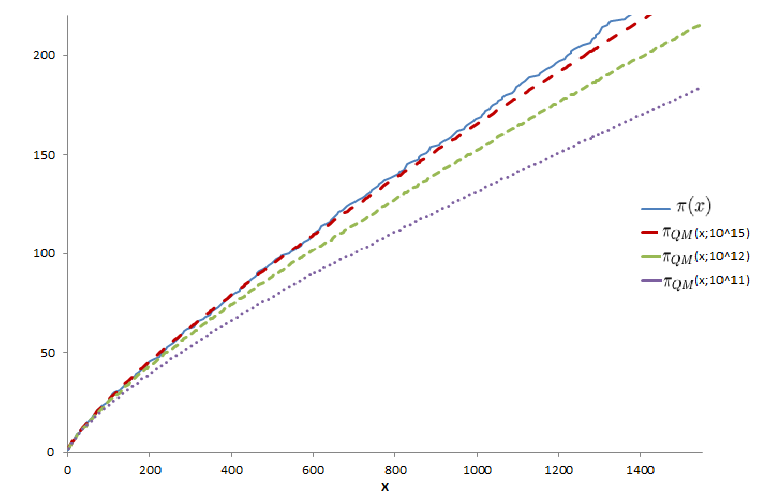}
\caption{$\pi_{QM}(x;N)$ vs actual $\pi(x)$}
\end{figure}

%\end{figure}

Remarkably, in the limit $N\rightarrow\infty$, $\kappa_{1}\rightarrow 0$, and we get $\lim_{j\rightarrow\infty} \mathcal{F}(j)=\Pi\{2\}$.
\begin{equation}
\pi_{QM}(x;N\gg 1)\simeq\pi(x).
\end{equation}\;

\section{Acknowledgment}
\label{sec:Acknowledgment}
This work has been partially  supported by Comunidad Aut\'{o}noma de Madrid,  project Quantum Information Technologies Madrid, QUITEMAD+ S2013-IC2801. I thank to Vicente Mart\'{i}n and Jes\'{u}s Mart\'{i}nez-Mateo for suggestions and assessment. \\ \\ \\

\end{document}